# Angular momentum of disc galaxies with a lognormal density distribution

John H Marr

*Unit of Computational Science, Building 250, Babraham Research Campus, Cambridge, CB22 3AT, UK.*

E-mail: john.marr@2from.com

**ABSTRACT**

By postulating that the majority of the mass and angular momentum of a disc galaxy is confined to the disc with a lognormal surface density distribution, and that galactic discs are substantially, if not fully, self-gravitating, it may be shown that the resultant rotation curves display a good overall fit to observational data for a wide range of galaxy types. With this hypothesis, the total angular momentum $J$ and total energy $|E|$ of 38 disc galaxies was computed and plotted against the derived disc masses, with best fit slopes for $J$ of 1.683±0.018 and $|E|$ of 1.643±0.038, and a universal disc spin parameter $\lambda$=0.423±0.014. Using the disc parameters $V_{max}$ and $R_{max}$ as surrogates for the virial velocity and radius, a virial mass estimator $M_{disc} \propto R_{max}V_{max}^2$ was generated, with a log-log slope of 1.024±0.014 for the 38 galaxies, and a proportionality constant $\lambda^* = 1.47 \pm 0.20 \times 10^5\ M_\odot kpc^{-1} km^{-2}\ s^2$. This relationship has less scatter than $M \propto (V_{rot})^\alpha$, and may provide an alternative to the Tully Fisher Relation in determining virial disc masses.

**Key words:** galaxies: fundamental parameters – galaxies: kinematics and dynamics – galaxies: individual: NGC 3198

## 1. INTRODUCTION

The classical picture of disc formation describes a well-mixed smoothly rotating halo, whose mass is determined by the virial theory, and with angular momentum (AM) induced through tidal torques. The assumption is then made that in-falling gas forms a disc, whose mass and AM are fixed fractions of the halo mass and AM. In this class of models, Hoyle (1949) and Peebles (1969) independently conjectured that galactic spins originated from tidal torques from neighbouring structures, and, more recently, Jones (1976) and Thuan & Gott (1977) showed that this mechanism can account to an order of magnitude for the values of galactic angular momenta.

An alternative solution to the problem of the origin of the AM of galaxies using cosmological turbulence models has been discussed by Thompson (1974) and Jones (1976), in which the AM is provided by the rotational motion of primordial turbulent eddies which were supposed to have escaped damping during the early phases of the history of the Universe, until they collapsed. Most of these theories on galaxy formation predict well-determined relationships between galaxy mass $M$, and AM $J$, which are not drastically different between these models: they are all found to be power laws of the form $J \propto M^n$, with n in the range 5/3–2 (Vettolani et al. 1980).

Models of galaxy formation and numerical cosmological simulations have both been used to estimate the size and rotation curves (RCs) of the discs, assuming the disc to be thin with an exponential surface density (Fall & Efstathiou 1980; Blumenthal et al. 1984; Mo, Mao & White 1998). It is generally assumed that dark matter (DM) and baryonic AM were initially the same, as both were subjected to identical early tidal torques. However, van den Bosch, Burkert & Swaters (2001) and Wise & Abel (2007) consider this unlikely because, although the two components experience the same torques and the same merging processes, they undergo very different relaxation, with heating and violent relaxation in DM halos, and shock heating plus feedback in the gaseous disc. The final state for the galaxy is compounded by radiation from the proto-halo and mergers to generate the observed galaxies, but it is probably fair to state that theoretical models are still uncertain, with simulations often predicting smaller discs than those found by observation (Navarro & Steinmetz 1997; Steinmetz & Navarro 1999; Navarro & Steinmetz 2000a; van den Bosch et al. 2001), and simultaneously accounting for the mass and AM distribution of spiral galaxies remains a major challenge for hierarchical formation models (Navarro & White 1994; van den Bosch 2002; Governato et al. 2007; Courteau et al. 2007). Although galaxy formation models and numerical simulations can mimic observations reasonably well, both are constrained by their initial assumptions, and are subject to evolution effects which may magnify any boundary errors over time, and both assume a large part of the AM to be retained by the halo.

## 2. MODELLING AM WITH A LOGNORMAL DENSITY DISTRIBUTION

In the absence of direct observation of an ethereal DM halo, an alternative approach is to take the observational assumption that the disc contains most of the mass and AM, including any gravitationally bound DM.

The assumption of a thin, axisymmetric disc with lognormal (LN) surface density distribution (equation 1) has been demonstrated to generate good fits to observational RCs for a wide range of galaxies (Marr 2015):

$$\Sigma(r) = \frac{\Sigma_0}{r/r_\mu} \exp\left(-\frac{\log^2(r/r_\mu)}{2\sigma^2}\right), \quad R_{max} \geq r > 0 \quad (1)$$

where $\Sigma(r)$ is the disc surface density ($M_\odot$ kpc$^{-2}$), $r$ is radial distance (kpc), $r_\mu$ is the mean of the natural logarithm of the radius (kpc), $\sigma$ is the standard deviation of the natural logarithm of the radius, $R_{max}$ is a maximum radius for the disc (kpc), and $\Sigma_0$ ($M_\odot$ kpc$^{-2}$) is a surface density parameter.

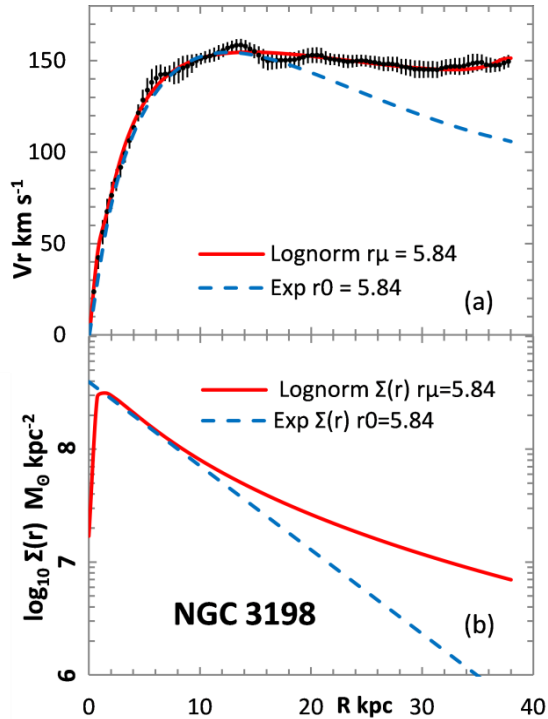

Fig 1. (a) RC for NGC 3198, with fitted LN (solid line) and exponential curve (dashed line). RC data de Blok et al. (2008). (b) Surface density plots for the two curves.

A typical RC is shown in Figure 1a for NGC 3198, with the fitted LN curve (red solid line). $R_{max}$ was taken to be 38 kpc, with values $\Sigma_0$=1.54x10$^8$ $M_\odot$/kpc$^2$, $r_\mu$=5.84 kpc, and $\sigma$ =1.20, and a small uniform bulge with radius 0.9 kpc and mass 1.01x10$^9$ $M_\odot$. The RC data and error bars were taken from the THINGS survey of de Blok et al. (2008), with a quoted recessional distance of 13.8 Mpc, compared to 9.65 Mpc in the original paper of van Albada et al. (1985), leading to a larger $R_{max}$ than that quoted in Marr's original paper, and an increase in modelled $M_{disc}$ of 7%. A freely-fitted exponential surface density curve is also shown for comparison (blue dashed line) with $r_0 = r_\mu$ ($r_0$ is the galaxy scalelength) and a free value for $\Sigma_0$ (4.00x10$^8$ $M_\odot$/kpc$^2$). The LN model gives a good fit throughout the RC with a fast rise at small radius and a long flattened tail at large $r$, while the exponential RC falls more rapidly at large $r$, with differing theoretical disc masses of 1.31x10$^{11}$ and 8.48x10$^{10}$ $M_\odot$ respectively.

The LN function has the property of generating a wide range of RCs with good agreement to the observational curves for a large selection of galaxy types, without requiring modified Newtonian dynamics (MOND) or a DM halo. The assumption of a LN surface density function also gives a plausible fit to a baryonic Tully Fisher Relation (bTFR), and enables other properties of the disc to be derived, such as its total theoretical AM and energy, enabling a spin parameter to be assigned for each galaxy.

### 2.1. The spin parameter λ

In analyses of AM, it is useful to define a dimensionless spin parameter, $\lambda$ (Peebles 1969; Navarro & Steinmetz 2000b; van den Bosch et al. 2001; Teklu et al. 2014):

$$\lambda \equiv \frac{J|E|^{1/2}}{GM^{5/2}} \quad (2)$$

For a thin exponential surface density disc of infinite extent, with total energy $E = K + PE$, and surface density $\Sigma(r) = \Sigma_0 \exp(-r/r_0)$, it may be shown analytically that the spin parameter is $\lambda \cong 0.425$ (Binney & Tremaine 2008). No comparable analytical function is available for the log normal surface density, but values for $J$ and $|E|$ can be computed for the function, and are listed for 38 representative galaxies in Table 1, with the corresponding computed spin parameter for each galaxy. Although some of the galaxies were assigned a small bulge to accommodate the RC at low $r$, as with NGC 3198, these carried none of the AM, and their masses were generally <1% of the disc mass. For the 38 galaxies, the mean value for the spin parameter is $\lambda$=0.423±0.014, suggesting that these LN model galaxies also have a universal spin parameter comparable to that of the exponential disc.

### 2.2. Obtaining a spin parameter for the LN disc

A log plot of $J$ and $|E|$ against the derived disc mass $M$ is shown in Figure 2, with rms best fit slopes of 1.683±0.018 and 1.643±0.038 respectively. Both parameters correlate tightly to a theoretical slope of 5/3, in conformity with the earlier work of Vettolani et al. (1980), and recent observational work from kinematic and photometric data that suggests $j^* \propto M^{*\alpha}$ and $\alpha = 0.6 \pm 0.1$, where $j^*$ is the specific angular momentum, $J^*/M^*$ (Romanowsky & Fall, 2012; Fall & Romanowsky 2013).

Assuming virial values for total mass, $M_{vir}$, and a velocity $V_{vir}$ at $R_{vir}$ (where $R_{vir}$ is a radius parameter for the disc) we may substitute these values into $J$ and $|E|$ to obtain equation 3:

$$M_{vir} = \frac{R_{vir}V_{vir}^2}{\sqrt{2}G\lambda} \quad (3)$$

In practice the virial parameters are unknown, but we do know the maximum radius $R_{max}$, and the peak velocity $V_{max}$, and equation 3 may be rewritten using these:

$$M_{disc} = \lambda^* R_{max} V_{max}^2 \quad (4)$$

where $\lambda^*$ is a constant for the disc system. A log plot of $R_{max}V_{max}^2$ against the model disc masses is shown in Figure 3 for the 38 galaxies, along with a classical TF plot using $V_{max}$. It may be noted that the $R_{max}V_{max}^2$ plots have less scatter than the $V_{max}$ plots, with slopes of 1.024±0.014 and 0.291±0.019 respectively, and corresponding intercepts of -5.422±0.150 and -0.961±0.198 (solid lines). Also shown in Figure 3 are the theoretical slopes for both plots, with slopes of 1 and 0.25 respectively. Again, the $R_{max}V_{max}^2$ plots lie much closer to the theoretical slope, suggesting a relationship:

$$M_{disc} = 1.47 \pm 0.20 \times 10^5 R_m V_m^2 \qquad (5)$$

where λ* has units of $M_\odot$ kpc$^{-1}$ (kms$^{-1}$)$^{-2}$.

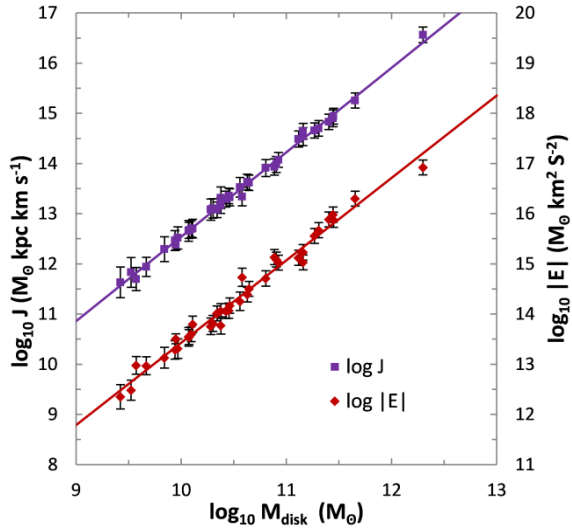

Fig 2. Log-log plots of $J$ and $|E|$ and the rms fits.

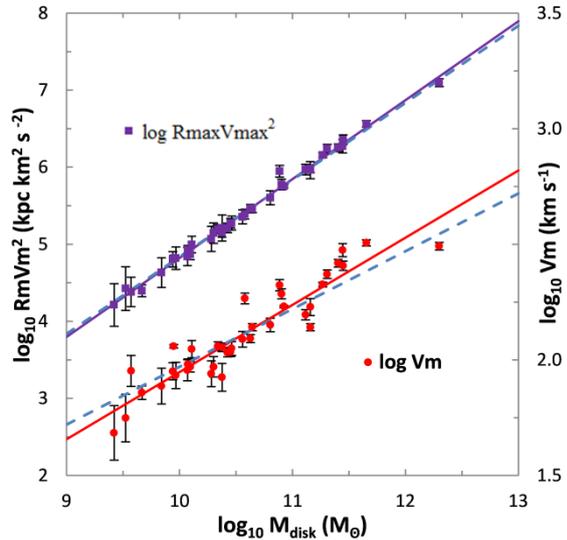

Fig 3. Log-log plots of $V$ and $RV^2$ with the rms fits (solid lines) and the theoretical slopes (dashed lines).

## 3. THE STABILITY OF THIN DISCS

One important justification for a DM halo is its stabilizing effect on thin discs. Gravitational stability of disc galaxies is generally differentiated into global stability and local stability. Analytical methods are not available for global instability, so this is generally checked using numerical simulations which suggest that, without a dark matter halo (or adding a very massive bulge/object in the centre), a disc galaxy is unstable to a bar-like configuration for a wide range of matter distribution, and it is unlikely that adding matter smoothly through the disc will stabilize it (Toomre 1981; Zasov, Khoperskov & Saburova, 2011).

Zasov however has stated that the lack of a DM halo is not a definite constraint on the stability of a thin disc, and in principle a disc of any density may be gravitationally stable; but the greater its density at a given $R$, the higher should be the velocity dispersion (and hence disc thickness) for the disc to be stable (personal communication 2015). This argument also applies to the MOND model, and further numerical analysis is needed to determine a minimum halo mass for stability with these models (Roshan & Abbassi 2015).

For local stability, Hunter and Toomre (1969) looked at the eigenmodes of warped galaxies, and were able to reduce the Laplace equation for the gravitational potential to a product of Legendre functions characterised by a single integer, *m*. They showed that a thin, self-gravitating disc in centrifugal equilibrium is stable to all vertical perturbations with *m=0* or *m=1*. For *m>1* there is no analytical proof of stability, but numerical normal mode analyses did not reveal any instabilities, and observed warps have *m=1* and so are definitely stable (Binney & Tremaine, 2008).

MOND predicts that the dynamically determined surface density of low surface-brightness galaxies will be much higher than the observed surface density (Milgrom 1989; Brada & Milgrom 1999), and when this surface density was used in calculating the Toomre-Q value, they generally found a much higher Q value, and hence greater stability, than under Newtonian dynamics. More recently, Roshan and Abbassi (2015) have derived the Toomre-like stability criterion for both fluid and stellar differentially rotating discs. Specifically, the stability criterion can be expressed in terms of a matter density threshold over which the instability occurs. They used a sample of six galaxies from the THINGS catalogue of spiral galaxies to compare MOND with Newtonian gravity, and investigated the possible and detectable differences between these theories. The rotation curve, epicycle frequency, and the velocity dispersion (*Vs*) are known for every galaxy in THINGS, enabling the stability parameter to be calculated with respect to *r* for both the stellar and gaseous components. For the gaseous component, they found *Vs* ~11 km s−1 and almost constant with respect to *r*, with the critical surface density for MOND smaller than for Newtonian gravity, but the scale on which this deviation will appear is at *r >~67 kpc*, whereas for most spiral galaxies the characteristic optical radius is <67 kpc. Therefore the difference between Newtonian dynamics and MOND might only be significant near the edge of the galactic disc, with some regions there that could be stable under Newtonian dynamics but unstable in MOND. However, they concluded it is unlikely that a galaxy will be found where the predictions of MOND are substantially different from those for Newtonian gravity (Roshan & Abbassi 2015).

## 4. DISCUSSION

Obreschkow & Glazebrook (2014) argue that mass and AM are the two most fundamental properties in galaxies, yet while most galaxy properties scale with galaxy mass, similar scaling relations for AM are only just being discovered. They add that, by combining observational data with analytical models and high-performance computer simulations, AM may become an essential tool of mainstream astronomy. The derivation of the *J/M* relationship of figure 2 is intermediate between purely theoretical models (e.g. Thompson 1974), and wholly observational curves (e.g. Fall 1983; Romanowsky & Fall 2012; Fall & Romanowsky 2013; Obreschkow & Glazebrook 2014). By postulating that the majority of the mass and angular momentum of a disc galaxy is confined to the disc with a LN surface density distribution, and that galactic discs are substantially, if not fully, self-gravitating, it has been shown that the resultant RCs display a good overall fit to the observational data for a wide variety of galaxy types and luminosities (Marr 2015), and the assumption of a universal LN surface density model allows theoretical disc masses and AM to be computed from purely observational data (the rotation curves).

**Table 1.** Galaxies modelled with a LN disc density distribution, associated parameters, observational peak rotation velocity, and theoretically derived total disc mass.

| Name | $\log_{10} \Sigma_0$ $M_\odot$ kpc$^{-2}$ | $r_\mu$ kpc | $\sigma$ | Rmax kpc | $\log_{10}$ Mdisc $M_\odot$ | Vpeak km s$^{-1}$ | $\log_{10}$ J $M_\odot$ kpc km s$^{-1}$ | $\log_{10}$ |E| $M_\odot$ km$^2$ s$^{-2}$ | $\lambda$ | Ref* |
|---|---|---|---|---|---|---|---|---|---|---|
| DDO 154 | 7.41 | 3.01 | 1.19 | 8.5 | 9.53 ± 0.15 | 56 ± 15 | 11.83 | 12.48 | 0.425 | 1,13 |
| F563-V2 | 8.26 | 2.35 | 1.22 | 12.0 | 10.34 ± 0.15 | 113 ± 5 | 13.08 | 14.01 | 0.417 | 2 |
| F568-1 | 8.19 | 3.90 | 1.25 | 15.0 | 10.65 ± 0.15 | 139 ± 5 | 13.62 | 14.50 | 0.415 | 2 |
| F568-3 | 7.88 | 5.00 | 1.18 | 14.0 | 10.43 ± 0.15 | 108 ± 5 | 13.29 | 14.06 | 0.415 | 2 |
| F568-V1 | 8.09 | 3.85 | 1.37 | 19.0 | 10.63 ± 0.15 | 124 ± 5 | 13.63 | 14.39 | 0.417 | 2 |
| F574-1 | 7.83 | 5.07 | 1.38 | 16.0 | 10.45 ± 0.15 | 108 ± 5 | 13.35 | 14.07 | 0.416 | 2,13 |
| M31 | 8.77 | 4.28 | 1.15 | 34.8 | 11.45 ± 0.15 | 255 ± 12 | 14.94 | 15.88 | 0.421 | 3 |
| Milky Way | 8.67 | 5.20 | 1.90 | 21.0 | 11.44 ± 0.15 | 298 ± 20 | 14.90 | 15.98 | 0.440 | 4,13 |
| NGC 925 | 7.66 | 10.14 | 1.39 | 15.0 | 10.56 ± 0.18 | 123 ± 10 | 13.53 | 14.25 | 0.417 | 1,13 |
| NGC 1705 | 7.65 | 4.10 | 1.70 | 4.8 | 9.67 ± 0.18 | 72 ± 5 | 11.94 | 12.97 | 0.417 | 5,13 |
| NGC 2403 | 8.19 | 4.29 | 1.40 | 20.0 | 10.80 ± 0.15 | 142 ± 10 | 13.91 | 14.71 | 0.416 | 1 |
| NGC 2683 | 9.12 | 1.26 | 1.27 | 20.0 | 10.89 ± 0.15 | 211 ± 12 | 13.93 | 15.13 | 0.444 | 6,13 |
| NGC 2841 | 9.27 | 2.10 | 1.70 | 35.5 | 11.65 ± 0.15 | 321 ± 10 | 15.25 | 16.30 | 0.431 | 1 |
| NGC 2903 | 8.83 | 2.50 | 1.55 | 32.0 | 11.27 ± 0.15 | 212 ± 5 | 14.66 | 15.56 | 0.427 | 1 |
| NGC 2915 | 7.96 | 2.83 | 1.63 | 16.0 | 10.30 ± 0.15 | 93 ± 10 | 13.10 | 13.81 | 0.418 | 5 |
| NGC 2976 | 7.98 | 2.50 | 1.60 | 3.0 | 9.57 ± 0.15 | 90 ± 15 | 11.70 | 12.98 | 0.416 | 1 |
| NGC 3198 | 8.19 | 5.84 | 1.20 | 38.0 | 11.12 ± 0.15 | 159 ± 4 | 14.49 | 15.12 | 0.419 | 1 |
| NGC 3521 | 8.92 | 2.50 | 1.40 | 31.5 | 11.31 ± 0.15 | 235 ± 10 | 14.70 | 15.67 | 0.428 | 1 |
| NGC 3726 | 8.22 | 5.94 | 1.40 | 32.0 | 11.16 ± 0.15 | 169 ± 15 | 14.54 | 15.23 | 0.417 | 7 |
| NGC 3741 | 7.44 | 2.52 | 1.56 | 7.0 | 9.42 ± 0.18 | 48 ± 15 | 11.63 | 12.35 | 0.415 | 8 |
| NGC 4217 | 8.68 | 2.97 | 0.97 | 16.0 | 10.90 ± 0.15 | 193 ± 10 | 13.98 | 15.09 | 0.426 | 7,13 |
| NGC 4389 | 7.98 | 3.70 | 1.28 | 5.0 | 9.95 ± 0.10 | 115 ± 2 | 12.38 | 13.51 | 0.421 | 7,13 |
| NGC 6946 | 8.46 | 3.67 | 1.13 | 19.5 | 10.92 ± 0.15 | 170 ± 2 | 14.07 | 15.02 | 0.435 | 1,13 |
| NGC 7331 | 8.97 | 2.79 | 1.45 | 26.0 | 11.40 ± 0.15 | 262 ± 10 | 14.83 | 15.88 | 0.423 | 1 |
| NGC 7793 | 8.24 | 2.57 | 0.67 | 8.0 | 10.11 ± 0.16 | 111 ± 10 | 12.70 | 13.79 | 0.494 | 1,13 |
| UGC 128 | 7.74 | 11.32 | 1.28 | 50.0 | 11.16 ± 0.15 | 138 ± 5 | 14.64 | 15.03 | 0.420 | 9,13 |
| UGC 2885 | 8.25 | 16.20 | 2.44 | 130.0 | 12.30 ± 0.15 | 310 ± 12 | 16.56 | 16.92 | 0.435 | 10,13 |
| UGC 5750 | 7.50 | 6.90 | 1.13 | 23.00 | 10.38 ± 0.15 | 84 ± 8 | 13.31 | 13.77 | 0.421 | 11,13 |
| UGC 6399 | 7.86 | 3.76 | 1.35 | 8.5 | 10.10 ± 0.15 | 93 ± 5 | 12.69 | 13.60 | 0.415 | 7 |
| UGC 6446 | 7.78 | 3.83 | 1.70 | 15.5 | 10.28 ± 0.15 | 87 ± 12 | 13.08 | 13.76 | 0.416 | 7 |
| UGC 6667 | 7.80 | 4.02 | 1.46 | 8.5 | 10.07 ± 0.18 | 90 ± 10 | 12.66 | 13.55 | 0.415 | 7 |
| UGC 6818 | 7.52 | 5.66 | 1.46 | 7.3 | 9.84 ± 0.18 | 77 ± 15 | 12.29 | 13.13 | 0.416 | 7 |
| UGC 6917 | 7.98 | 4.17 | 1.46 | 11.0 | 10.38 ± 0.15 | 113 ± 5 | 13.16 | 14.06 | 0.415 | 7 |
| UGC 6923 | 7.83 | 3.55 | 1.55 | 9.0 | 10.07 ± 0.15 | 96 ± 5 | 12.67 | 13.54 | 0.415 | 12 |
| UGC 6969 | 7.56 | 5.94 | 1.39 | 8.0 | 9.94 ± 0.18 | 89 ± 8 | 12.46 | 13.29 | 0.416 | 12 |
| UGC 6973 | 9.07 | 0.96 | 1.76 | 7.3 | 10.58 ± 0.18 | 184 ± 10 | 13.34 | 14.73 | 0.420 | 7 |
| UGC 6983 | 8.16 | 3.06 | 1.21 | 15.0 | 10.46 ± 0.15 | 112 ± 8 | 13.32 | 14.17 | 0.416 | 7 |
| UGC 7089 | 7.60 | 4.75 | 1.50 | 9.0 | 9.97 ± 0.18 | 86 ± 12 | 12.52 | 13.31 | 0.415 | 7 |

*Sources for Rmax, Vpeak and Verror: (1) de Blok et al. (2008); (2) Swaters, Madore & Trewhella (2000); (3) Carignan et al. (2006), Rubin & Kent Ford (1970); (4) Bhattacharjee, Chaudhury & Kundu (2014); (5) Elson, de Blok & Kraan-Korteweg (2012); (6) Casertano & van Gorkom (1991); (7) Sanders & Verheijen (1998); (8) Begum, Chengalur & Karachentsev (2005); (9) de Blok & McGaugh (1998); (10) Roelfsema & Allen (1985); (11) de Blok, McGaugh, & Rubin (2001); (12) Bottema (2002); (13) these included a bulge in the original RC calculation (Marr 2015) (see text).

The resultant $J/M$ log plots have the same slope of 5/3 as the observational plots of Romanowsky and Fall (2012), but with less scatter. This may reflect the variety of galaxy morphologies of the 67 Sa–Sm spirals selected by Romanowsky and Fall which consequently have differing bulge to mass ratios reflecting their classification, whereas the LN plots are based solely on the rotation curves and their theoretically derived disc masses. At the limit of ellipticals, Romanowsky and Fall found that these too have a $J^*/M^*$ slope of 5/3, but are shifted in $J^*$ by a factor

~5, or in mass by a factor ~11, presumably reflecting the fact that most of the stars in ellipticals are not rotating about the galactic centre. This is taken further by Obreschkow and Glazebrook (2014) who analysed 16 disc galaxies from the THINGS survey to present detailed triaxial plots of baryonic masses, angular momenta and bulge mass fractions ($\beta$). They found a close correlation in their plots with $\beta$ as a dependent variable, but again these were consistent with an angular momentum/disc mass slope of 5/3. Obreschkow and Glazebrook suggest that the lower $J/M$ values in ellipticals reflect a significant loss in angular momentum in their formation history. However, an alternative interpretation may be that their present observed values reflect conservation of their formation values, with initial $J/M$ ratios insufficient to allow disc formation to occur.

A number of theoretical predictions propose the existence of DM within galaxies to flatten the rotation curve, to stabilise the disc, and to address mass discrepancy problems in the Universe. Although unable to accommodate these measures of DM, the theoretical total disc masses generated by a LN density distribution model can accommodate a scenario in which the total mass distribution is confined to the disc to account for any remaining discrepancy between the observational and theoretical disc mass. Pfenniger & Revaz (2005) have suggested that the baryonic disc mass is likely to contain a dark baryonic component proportional to the HI gas in addition to the detected baryons, stars and gas, and the bTFR can be substantially improved when the HI mass is multiplied by a factor of about 3, while Maller & Bullock (2004) and Fukugita & Peebles (2006) have also suggested that ionized (warm) gas in the more massive galaxies may be more significant in this respect. Gurovich et al. (2010) and Jalocha et al. (2010) considered that, at smaller scales, the contribution of non-baryonic DM to spiral galaxy masses could be much less than anticipated in spherical halo models, and a larger fraction of undetected baryons may be required in the more massive galaxies to steepen the slope of the theoretical bTFR to its observed value. This reinforces the suggestion made by several workers that mass within galactic discs must be a multiple of the HI mass, and that galactic discs may be substantially, if not fully, self-gravitating (Bosma 1981; Hoekstra, van Albada & Sancisi 2001). Nevertheless, irrespective of its formation history or of the presence of a DM halo, the disc contributes the major part of the total light and the observable AM. In M31 for example, >95% of the total AM and >75% of the blue light come from its disc (de Vaucouleurs 1958; Takase 1967).

The theoretical instability of thin discs presents a valid argument against them, and the LN disc may suffer from the same instability criteria as the MOND disc in the absence of a DM halo. This has been examined for MOND by Roshan and Abbassi (2015), but has not yet been resolved for a LN disc. Whilst MOND has gone some way towards explaining RCs in the absence of a DM halo, doubt remains about its physical justification, and mathematically it can equally explain the flattened RCs by assuming attenuation of the inertial mass rather than augmentation of the gravitational field in the weak field limit.

In their classic paper, Tully & Fisher (1977) presented both the luminosity/global profile width and a mass/diameter-profile width squared relation to determine their distances to field galaxies, and assumed $M/L \cong constant$ to give equal weight to both luminosity plots. Van den Bosch (2002) modelled disc formation from DM halos by accretion over time, and suggested that the luminosities and circular velocities of disc galaxies were poor indicators of total virial mass in these models, and showed that the product of disc scalelength and rotation velocity squared (the virial mass estimator) yields a much more robust estimate of virial mass. In the LN model, log plots of $M_{disc} \propto R_{max}V_{max}^2$ show less scatter than the standard plot of $M \propto V^\alpha$, with a slope close to its theoretical value of 1 for a wide variety of galaxy types and luminosities.

The assumption of a LN density distribution enables a good overall fit to the observational RC data using one well-defined model (Marr 2015). The tight correlation of AM with the total gravitational disc mass suggests a fundamental relationship, and may provide a rationale for the bTFR. The derived angular momenta and total energies for the LN discs correlate well to theoretical models, with a plausible value for a universal disc spin parameter λ* as the basis for a virial mass estimator. Because the velocity line widths and disc radii are relatively easy to determine, the virial mass estimator of radius times velocity squared may provide an alternative estimator for galactic disc masses.

## ACKNOWLEDGMENTS

I would like to thank Erwin de Blok for supplying the detailed data for NGC 3198, Mahmood Rashin and Anatoly Zasov for valuable discussions about disc stability, and the anonymous referee for many helpful suggestions and comments.

## REFERENCES

Begum A., Chengalur J. N., Karachentsev I. D., 2005, A&A, 433, L1
Bhattacharjee P., Chaudhury S., Kundu S., 2014, ApJ, 785, 63
Binney J., Tremaine S., 2008, in: Galactic Dynamics, Princeton Univ.Press, Princeton, NJ
Blumenthal G.R., Faber S.M., Primack J.R., Rees M.J., 1984, Nature, 311, 517
Bosma A., 1981, AJ, 86, 1825
Bottema R., 2002, A&A, 388, 793
Brada R., Milgrom M., 1999, ApJ, 519, 590
Carignan C., Chemin L., Huchtmeier W.K., Lockman F. J., 2006, ApJ, 641, L109
Casertano S., van Gorkom J. H., 1991, AJ, 101, 1231
Courteau S., Dutton A. A., van den Bosch F. C., MacArthur L. A., Dekel A., McIntosh D. H., Dale D. A., 2007, ApJ, 671, 203
de Blok W. J. G., McGaugh S.S., 1998, ApJ, 508, 132
de Blok W. J. G., Walter F., Brinks E., Trachternach C., Oh S-H., Kennicutt R. C., 2008, AJ, 136, 2648
de Blok W.J.G., McGaugh S.S., Rubin V.C., 2001, AJ, 122, 2396
de Vaucouleurs G., 1958, ApJ, 128, 465
Elson E. C., de Blok W. J. G., Kraan-Korteweg R. C., 2012, AJ, 143, 1
Fall M., Efstathiou G., 1980, MNRAS, 193, 189
Fall S. M., 1983, in Athanassoula E., ed., Proc. IAU Symp. 100, Internal Kinematics and Dynamics of Galaxies. Cambridge Univ. Press, Cambridge, p. 391 (F83)
Fall S. M., Romanowsky A.J., 2013, ApJ Letters, 769, L26
Fukugita M., Peebles P. J. E., 2006, ApJ, 639, 590
Governato F., Willman B., Mayer L., Brooks A., Stinson G., Valenzuela O., Wadsley J., Quinn T., 2007, MNRAS, 374, 1479

Gurovich S., Freeman K., Jerjen H., Staveley-Smith L., Puerari I., 2010, AJ, 140, 663

Hoekstra H., van Albada T. S., Sancisi, R., 2001, MNRAS, 323, 453

Hoyle F., 1949, in Proc. of the Symp. on the Motion of gaseous Masses of Cosmical Dimensions, Problems of Cosmical Aerodynamics. ed. J. M. Burgers & H. C. Van de Hulst. Central Air Documents Office, Dayton, p. 195

Hunter C., Toomre A., 1969, ApJ, 155, 747

Jalocha J., Bratek L., Kutschera M., Skindzier P., 2010, MNRAS, 406, 2805

Jones B. J. T., 1976, Rev. Mod. Phys., 48, 107

Maller A. H., Bullock J. S., 2004, MNRAS, 355, 694

Marr J.H., 2015, MNRAS, 448, 3229

Milgrom M., 1989, ApJ, 338, 121

Mo H. J.; Mao S., White S. D. M., 1998, MNRAS, 295, 319

Navarro J. F., Steinmetz M., 1997, ApJ, 478, 13

Navarro J.F., Steinmetz M., 2000a, ApJ, 528, 607

Navarro J.F., Steinmetz M., 2000b, AJ, 538, 477

Navarro J. F., White S. D. M., 1994, MNRAS, 267, 401

Obreschkow D., Glazebrook K., 2014, ApJ, 784, 26

Peebles, P. J. E., 1969, ApJ, 155, 393

Pfenniger D., Revaz Y., 2005, A&A, 431, 511

Roelfsema P. R., Allen R. J., 1985, A&A, 146, 213

Romanowsky, A. J., Fall, S. M., 2012, ApJS, 203, 17

Roshan M., Abbassi S., 2015, ApJ, 802, 9

Rubin V. C., Kent Ford W., 1970, ApJ, 159, 379

Sanders R. H., Verheijen M. A. W., 1998, ApJ, 503, 97

Steinmetz M., Navarro J.F., 1999, ApJ, 513, 555

Swaters R. A., Madore B. F., Trewhella M., 2000, ApJ, 531, L107

Takase B., 1967, PASJ, 19, 427

Teklu A., Remus R.-S., Dolag K., Burkert A., 2014, in Ziegler B. L., Combes F., Dannerbauer H., Verdugo M., eds, Proc. IAU Symp. 309, Galaxies in 3D across the Universe. Cambridge Univ. Press, Cambridge, p. 349

Thompson L. A., 1974, PhD Thesis, Univ.Arizona.

Thuan T. X., Gott J. R., III, 1977., ApJ, 216,194

Toomre A., 1981, in Fall M., Lynden-Bell D., eds,The Structure and Evolution of Normal Galaxies. Cambridge Univ. Press, p. 111

Tully R. B., Fisher J. R., 1977, A&A, 54, 661

van Albada T. S., Bahcall J. N., Begeman K., Sancisi R., 1985, Ap J, 295, 305

van den Bosch F.C., 2002, MNRAS, 332, 456

van den Bosch F.C., Burkert A., Swaters R.A., 2001, MNRAS, 326, 1205

Vettolani G., Marano B., Zamorani G., Bergamini R., 1980, MNRAS, 193, 269

Wise J. H., Abel T., 2007, ApJ, 665, 899

Zasov A.V., Khoperskov A.V., Saburova A.S., 2011, Astron. Lett., 37, 374